# Muon-spin-relaxation study of the magnetic penetration depth in MgB$_2$

Ch. Niedermayer[1], C. Bernhard[2], T. Holden[2], R.K. Kremer[2] and K. Ahn[2]

[1] Universität Konstanz, Fachbereich Physik, D-78457 Konstanz, Germany

[2] Max-Planck-Institut für Festkörperforschung, Heisenbergstrasse 1, D-70569 Stuttgart, Germany

The magnetic vortex lattice (VL) of polycrystalline MgB$_2$ has been investigated by transverse-field muon-spin-relaxation (TF-μSR). The evolution of TF-μSR depolarization rate, $\sigma$, that is proportional to the second moment of the field distribution of the VL has been studied as a function of temperature and applied magnetic field. The low temperature value $\sigma$ exhibits a pronounced peak near $H_{ext}$ = 75 mT. This behavior is characteristic of strong pinning induced distortions of the VL which put into question the interpretation of the low-field TF-μSR data in terms of the magnetic penetration depth $\lambda(T)$. An approximately constant value of $\sigma$, such as expected for an ideal VL in the London-limit, is observed at higher fields of $H_{ext}$ > 0.4 T. The TF-μSR data at $H_{ext}$ = 0.6 T are analyzed in terms of a two-gap model. We obtain values for the gap size of $\Delta_1$ = 6.0 meV ($2\Delta_1/k_BT_c$ = 3.6), $\Delta_2$ = 2.6 meV ($2\Delta_2/k_BT_c$ = 1.6), a comparable spectral weight of the two bands and a zero temperature value for the magnetic penetration depth of $\lambda_{ab}$ ≈ 100 nm. In addition, we performed μSR-measurements in zero external field (ZF-μSR). We obtain evidence that the muon site (at low temperature) is located on a ring surrounding the center of the boron hexagon. Muon diffusion sets in already at rather low temperature of T > 10 K. The nuclear magnetic moments can account for the observed relaxation rate and no evidence for electronic magnetic moments has been obtained.

The recent discovery of superconductivity at 39 K in the binary compound $MgB_2$ [1] has triggered an enormous scientific effort in order to understand the mechanism that leads to such a high critical temperature in a seemingly classical superconductor. Based on band structure calculations Kortus et al. [2] conclude that a sizeable electron phonon coupling in combination with the high phonon frequencies due to the light mass of boron can reproduce the high critical transition temperature. The observation of a large B isotope effect confirms the important role of the phonons for superconductivity in this compound [3,4].

Subsequently, a large body of experimental work focused on the study of the symmetry of the superconducting order parameter. Measurements of the $^{11}B$ nuclear spin lattice relaxation rate [5], inelastic neutron scattering measurements [6], specific heat [7], high resolution photoemission [8] and scanning tunneling spectroscopy [9] are strongly in favor of a conventional BCS s-wave pairing state with moderately strong electron phonon coupling. Even if superconductivity is phonon mediated an analysis beyond a simple isotropic model may be required. Two different order parameters arising from two different sheets of the Fermi surface were calculated by Liu et al. [10] and these predictions are in good agreement with the analysis of specific heat [11] and Raman [12] data in terms of a two-gap model.

Recent muon-spin-rotation (μSR) and low field ac-susceptibility measurements have been interpreted in terms of a quadratic temperature dependence of the magnetic penetration depth. From their data, the authors concluded that $MgB_2$ is an unconventional superconductor with an energy gap that has nodes at certain points in k-space [13]. These first μSR experiments have been performed in a comparably small external magnetic field of 45 mT. At such a low magnetic field the VL is rather soft and therefore can easily be disordered by pinning induced distortions. In such a case random distortions of the VL will lead to a significant increase of the second moment of the magnetic field distribution, such as probed by the TF-μSR technique. As a result, the magnetic penetration depth cannot be reliably deduced from the TF-μSR data, at least not in a straightforward manner.

Here we present a more extensive set of TF-μSR data that spans a wider range of applied magnetic fields. We present evidence that pinning induced distortions of the vortex lattice are important at low magnetic fields for $H_{ext} \leq 0.3$ T, whereas they become less important at higher fields. We analyze the temperature dependence of the TF-μSR relaxation rate at 0.6 T in terms of the two-gap model. Using this model we obtain a good fit to our experimental data. The obtained parameters for the gap sizes, the relative density of states of the bands, and the absolute value of the magnetic penetration depth are in reasonable agreement with values that have been previously reported (see for example table I in [14]).

A polycrystalline $MgB_2$ sample has been prepared from a stoichiometric mixture of Mg (99.98 %, Johnson Matthey) and natural Boron powder (~ 60 mesh, 99.5 %, Aldrich). The reaction was carried out at 850 °C for two days using sealed Ta capsules under Ar atmosphere that were in turn encased in evacuated silica ampoules. After grinding under argon atmosphere, the sample was pressed into a pellet of 10 mm diameter using a cold isostatic press under the pressure of 0.3 GPa. The pressed pellet was annealed at the same temperature for one day. The x-ray powder diffraction pattern is characteristic of $MgB_2$ with lattice parameters of a=3.08529(8) Å and c=3.52384(7) Å. No indication for additional impurity phases was found. DC-susceptibility measurements using a commercial SQUID magnetometer reveal a sharp superconducting transition at $T_c = 38.2(1)$ K.

The muon spin rotation (μSR) experiments have been performed at the πM3 muon beam line at the Paul-Scherrer-Institut in Villigen, Switzerland. Disc shaped pellets of $MgB_2$ (10 mm in diameter and about 2 mm thick) were cooled in an externally applied magnetic field $B_{ext}$ to temperatures below $T_c$ in order to introduce a homogeneous flux line lattice (FLL). 100% spin-polarized positive muons are then implanted into the bulk of the sample with the initial muon spin polarization transverse to the external field. The muons thermalize rapidly without any significant loss in polarization and come to rest about 100 –200 μm below the surface typically at interstitial lattice sites. The implanted muons are randomly distributed throughout the field profile of the FLL since the magnetic penetration depth λ significantly exceeds the lattice constants. Each muon starts to process in its local magnetic field $B_{loc}(r)$ with the Larmor frequency $\omega_\mu = \gamma_\mu B_{loc}(r)$ where $\gamma_\mu$ = 851.4 MHz / T is the gyromagnetic ratio of the muon. The time evolution of the so-called muon spin polarization function $P_\mu(t)$ is measured by monitoring the decay positrons which are preferentially emitted along the muon spin direction at the instant of decay (half life time 2.2 μs). $P_\mu(t)$ is oscillatory in character with a rapidly decreasing amplitude. Under certain conditions (see discussion below), the depolarization of the muon spin precession provides a measure of the field distribution within the vortex state and hence of the magnetic penetration depth $\lambda$.

For polycrystalline samples the depolarization is approximately of Gaussian form, $P_\mu(t) \propto \exp(-\sigma^2 t^2 / 2)$, where the depolarization rate $\sigma$ is proportional to the second moment of the field distribution $\sigma \propto \langle \Delta B^2 \rangle^{1/2}$. For an isotropic type II superconductor the second moment $\langle \Delta B^2 \rangle$ is directly related to the magnetic penetration depth $\lambda$

$$\langle \Delta B^2 \rangle_{iso} = 0.0371 \; \Phi_0^2 \; \lambda^{-4} \text{ and thus } \sigma \left[ \mu s^{-1} \right] = 7.904 \times 10^4 \times \lambda^{-2} \left[ nm \right]$$

However, recent measurements of the upper critical field on single crystals [15], -c-axis oriented thin films [16] and aligned $MgB_2$ crystallites [17] give evidence for a sizeable anisotropy of the superconducting properties with anisotropy ratios ranging from about 1.6 to about 2.5. In this case both components of the magnetic penetration depth affect the value of the relaxation rate. The value of the anisotropy therefore needs to be known in order to deduce the components of the magnetic penetration depth. For the special case of a large anisotropy with $\gamma = \lambda_c / \lambda_{ab} > 5$ Barford and Gun [18] have shown that the measured effective penetration depth $\lambda_{eff}$ is independent of the anisotropy ratio since it is solely determined by the in plane penetration depth $\lambda_{ab}$

$$\lambda_{eff} = f_{anisotropy} \cdot \lambda_{ab} \text{ with } f_{anisotropy} = 1.23$$

$$\sigma \left[ \mu s^{-1} \right] = 7.086 \times 10^4 \times \lambda_{ab}^{-2} \left[ nm \right].$$

For anisotropies between 1 and 5 the correction factor $f_{anisotropy}$ varies between 1 and 1.23. For $\gamma = 2.5$, $f_{anisotropy} = 1.19$, which is already close to the high anisotropy limit.

It should be stressed that the above equations are valid only for the case of an ideal FLL and in the so-called London-limit for $\kappa = \lambda / \xi \gg 1$. With $\xi_{ab} \approx 7$ nm [17] and $\lambda_{ab} \approx 100$ nm the London-limit is applicable for $MgB_2$. Another point of concern, that needs to be addressed, is the question as to how much pinning effects introduce distortions of the VL. It has been shown theoretically and experimentally that random distortions of the VL can significantly enhance the μSR relaxation rate and thus lead to a sizeable underestimation of the value of the magnetic penetration depth [19]. Furthermore, the influence of the pinning effects will be temperature dependent and therefore can easily account for an unconventional temperature dependence of the μSR relaxation rate in the presence of a conventional SC order parameter. The influence of pinning on the μSR relaxation rate can be tested via the dependence of the μSR relaxation rate on the applied magnetic field. The London-model predicts that the second moment of the magnetic field distribution of a perfect VL should be independent of $H_{ext}$. The distortions of the VL due to pinning will be largest for small external fields where the vortices are far apart and their mutual interaction is comparably weak. At higher fields the vortex-vortex interaction is enhanced and it is more likely to maintain the long-range order of the vortex lattice. A prominent example are the cuprate high-$T_c$ superconductors, where one typically observes a peak in σ(H) at low field that is followed by a plateau at high field where σ($H_{ext}$) is almost constant (see e. g. Ref. [20] for $YBa_2Cu_3O_7$ and Ref. [21] for $Tl_2Ba_2CuO_{6+\delta}$).

Figure 1 shows the low-temperature μSR relaxation rate, σ(H, T = 5 K) as a function of the applied magnetic field for 10 mT < $H_{ext}$ < 0.64 T (the largest available TF- field of the spectrometer at PSI). Each point has been obtained by field-cooling the sample from above $T_c$ to 5 K. The value of σ(H,5 K) increases almost linearly below 50 mT, it goes through a pronounced maximum around 75 mT before it rapidly decreases again from 20 μs$^{-1}$ at the peak position to about 8 μs$^{-1}$ at 0.64 T. As outlined above, the observed peak in σ(H) is characteristic of pinning induced disorder of the FLL. This finding implies that the μSR relaxation rate obtained at such low magnetic field cannot be interpreted in terms of the magnetic penetration depth, at least not in a straightforward manner. Only at high magnetic fields for $H_{ext} \geq 0.4$ T the value of σ(H,5 K) is almost independent of $H_{ext}$ such as is expected for an ideal FLL in the London-limit. Figure 2 shows the evolution of the TF-μSR depolarization rate as a function of temperature for different external fields of $H_{ext}$ = 0.05 T, 0.3 T and 0.6 T. Shown by the solid lines are the best fits obtained with the function σ(T) = σ(0)·(1-(T/$T_c$)$^\alpha$) which is based on the empirical two-fluid model. We ascribe no particular physical meaning to this function, we simply intended to parameterize our data in order to describe the changes in the T-dependence of σ($H_{ext}$,T) with the external magnetic field. The T-dependence can be seen to change appreciably as a function of applied field with α = 1.33(8) at 50 mT, α = 1.44 (7) at 0.3 T and

α = 1.92(5) at 0.6 T. The T-dependence of σ(T) at low T therefore weakens as the applied magnetic field is increased.

We have outlined above, that the TF-μSR data obtained at $H_{ext}$ = 0.6 T can be most reliably interpreted in terms of the magnetic penetration depth. Figure 3 displays the temperature dependent TF-μSR depolarization rate at $H_{ext}$ = 0.6 T. The inset shows a magnification of the behavior at low temperature. It occurs that σ(T) is almost temperature independent below 5 K, while it exhibits a kink around 7 K where it suddenly begins to decrease with increasing temperature. Such a trend can be rather well reproduced with a two-gap model [14] that assumes that the SC carriers reside in two different energy bands. The coupling between these bands needs to be sufficiently weak such that the magnitude of the energy gaps is different while they both appear simultaneously at the $T_c$ that is determined by the larger gap. Evidence in favor of the two-gap model has been recently obtained by specific heat [11] and by Raman measurements [12]. The thick solid line shows the best fit to our experimental data using the two-gap model. The T-dependence of the SC condensate density $n_s(T) \sim \sigma(T)$ is assumed to be:

$$n_s(T) = n_s^0 - \gamma \cdot \delta n_s(\Delta_1, T) - (1-\gamma) \cdot \delta n_s(\Delta_2, T)$$

$$\delta n_s(\Delta, T) = \frac{2 n_s^0}{k_B T} \int_0^\infty f(\varepsilon, T)(1 - f(\varepsilon, T)) d\varepsilon$$

The parameter γ determines the ratio between the density of states of the band with the larger gap with respect to the one with the smaller gap, $k_B$ is the Boltzmann constant, and $f(\varepsilon, T)$ is the Fermi distribution of the quasi-particles with ε the energy of the normal electrons relative to the Fermi-energy.

$$f(\varepsilon, T) = \left(1 + e^{\frac{\sqrt{\varepsilon^2 - \Delta^2(T)}}{k_B T}}\right)^{-1}$$

For Δ(T), we used the BCS values tabulated by Mühlschlegel et al [22]. The obtained values of the fitting parameters are $\Delta_1$ = 6.0 meV ($2\Delta_1/k_B T_c$ = 3.6), $\Delta_2$ = 2.6 meV ($2\Delta_2/k_B T_c$ = 1.6) and γ=1.5 - 2 (the fit is not very sensitive to γ). These values are in reasonable agreement with previous results that have been obtained by applying the two-gap model to experimental data (see e.g. Bouquet et al. [14] and references therein). Taking the fitted low temperature value for $\sigma(T \rightarrow 0) = 7.9$ µs$^{-1}$ we derive $\lambda_{eff} = 100$ nm and, depending only weakly on the anisotropy, $\lambda_{ab}$ values between 95 nm and 100 nm, in good agreement with the value of $\lambda_{ab} = 110$ nm, reported by Manzano and Carrington [23]. In the previous µSR experiment which was performed at a rather low external field of $H_{ext} = 450$ Oe a value of $\lambda_{ab} = 85$ nm was obtained which is also rather close to our value. Apparently, the value of $\sigma(5 K) = 10$ µs$^{-1}$ obtained at $H_{ext} = 450$ Oe by Panagopoulos et al [13] is significantly lower than our value of $\sigma(5K) = 18$ µs$^{-1}$ at $H_{ext} = 500$ Oe. Such a difference could be explained if the pinning was much weaker in the commercial Alfa Aesar sample used by Panagopoulos et al. Another reason may be that the condensate density is significantly reduced in the commercial sample since it is known to contain a certain amount of magnetic impurities. Future experiments of the temperature and the field dependence of the TF-µSR depolarization rate on samples of different purity will be required in order to answer this question.

Let us return to the interpretation of the T-dependence of the TF-µSR depolarization rate. Irrespective of the good agreement with the two-gap model, one should keep in mind that a reasonable fit to the experimental data can be obtained also assuming other kind of scenarios, for example assuming that a single energy gap exists which is strongly anisotropic in k-space. In that case the temperature dependence of the magnetic penetration depth would be strongly modified by the presence of impurities. For a conventional anisotropic s-wave gap the impurity scattering would tend to reduce the anisotropy since it mixes different states in k-space. The clean-limit scenario of an anisotropic conventional s-wave gap has recently been discussed by Haas and Maki [24] who treated the case that the gap is significantly larger in the direction perpendicular to the boron planes than in the parallel one. It was shown earlier by Schneider and Singer that a sizeable anisotropy of the SC energy gap with value of $\Delta_{min}/\Delta_{max}$ of the order of 0.5-0.6 will be required in order to account for a $T^2$-dependence of $\sigma(T)$ [25]. In the case of an unconventional order parameter that changes sign in k-space (for example the $d_{x^2-y^2}$-wave OP in high $T_c$ cuprates), the potential scattering would tend to enlarge the nodal regions where the energy gap is zero [26]. As a result the absolute value of the magnetic penetration depth would be strongly reduced and the temperature dependence of $\lambda$ would tend to change from a linear T-dependence in the clean limit to a $T^2$-dependence in the dirty limit. Such a scenario has been proposed in Ref. [13] in order to explain the T-dependence of the TF-μSR relaxation rate. Shown by a dashed line is the result of a $T^2$-fit to our μSR data at 0.6 T. Evidently, this model also allows one to obtain a reasonable fit to the experimental data. For comparison the dotted line also shows the best fit assuming an isotropic single gap which does not describe our experimental data very well. The inset of Fig. 3 shows the most relevant low-temperature range on an enlarged scale. Such a closer inspection of the low-T data suggests that the two-gap model does provide the best fit to the experimental data.

Nevertheless, we would like to emphasize that based on experimental TF-μSR data one has to be very cautious with the attempt to discriminate between different scenarios concerning the nature and the symmetry of the superconducting energy gap. The experimental error of the data points, the problems related to pinning induced disorder of the flux-line lattice, possible structural changes of the vortex configuration as a function of temperature and the unknown effect of vortex dynamics, especially at large external field and in the vicinity of the SC transition, make it virtually impossible to resolve the rather small differences in the T-dependence of the magnetic penetration depth that are expected between the various scenarios. For example, we do presently not know what is the origin of the sudden change of σ(T) around 15 K. The most likely reason is a transition of the VL which concerns either the static order or else the vortex dynamics. Further μSR experiments, preferably on single crystalline materials will be required in order to shed more light on the static and dynamic behavior of the VL in $MgB_2$. We would like to stress that in our opinion only μSR experiments on single crystalline materials of good quality will allow one to obtain reliable μSR results concerning the temperature dependence of λ(T). This is in fact a lesson that was learned from studying the magnetic penetration depth of cuprate high $T_c$ superconductors. Experiments on polycrystalline materials as well as on the first available single crystals seemed to indicate that the magnetic penetration depth has a very weak low-T dependence that is indicative of an isotropic s-wave order parameter [20,27]. Only after the microwave measurements of Hardy and coworkers [28], it was recognized from studies on good single crystalline materials that the magnetic penetration depth rather follows a linear low-T dependence characteristic of an order parameter that has nodes in k-space. Meanwhile it is well established that the order parameter has mainly $d_{x^2-y^2}$ symmetry. It is still not understood why the T-dependence of the TF-µSR depolarization rate is so different for polycrystalline and single crystalline materials.

Finally, in order to check our samples for the presence of magnetic impurity phases, we performed zero field µSR measurements. Typical µSR time spectra are shown in Fig. 4. The static and randomly oriented nuclear magnetic moments of Boron (and to a lesser extent Magnesium) give rise to a Kubo-Toyabe like relaxation, which is characterized by a Gaussian decay at early times followed by a recovery of the asymmetry to 1/3

$$P_z(t) = \frac{1}{3} + \frac{2}{3}\left(1 - \Delta^2 t^2\right)\exp\left(-\frac{1}{2}\Delta^2 t^2\right) \qquad \text{(static)}$$

where $\Delta$ is the rms width of the field distribution arising from the nuclear moments. Besides the signature of these static nuclear magnetic moments, we obtained no evidence for any kind of additional magnetic moments (neither static nor slowly fluctuation within the characteristic time-window of the µSR experiment of $10^{-6} < \tau < 10^{-9}$ s). The spectrum obtained at 1.8K shows a recovery of the polarization to 1/3 in the time range shown. The spectrum obtained at 12.5 K shows a significant suppression of the 1/3 tail, whereas the short time behavior of $P_z(t)$ is not much changed from the static case. This behavior is characteristic for a slow motion of the muon with hop rates less than $\Delta^{-1}$. Best fit results were obtained by using a dynamic Kubo-Toyabe function of the form [29]

$$P(t) = P_z(t)\exp(-\nu t) + \nu \int_0^t P_z(t')\cdot \exp(-\nu t')\cdot P_z(t-t')dt' \qquad \text{(dynamic)}$$

where $\nu = 1/\tau_{hop}$ is the jump rate ($\tau_{hop}$ is the time between jumps in the diffusion process, i.e. the residence time in each potential well). A series of experiments done with a special setup which allows extension of the time range up to 20 µs shows that even at low temperatures the muon is not completely static, but shows very slow dynamics with hop rates of the order of 0.1 µs$^{-1}$. At 60 K the time evolution of the muon polarization is well described by an exponential decay $P_z(t) = \exp(-2\Delta^2 \tau_c t)$, indicating sufficiently fast motion of the muon ($\Delta\tau_c \ll 1$). Fast muon diffusion could in principal also lead to a motional narrowing of the field distribution arising from the flux line lattice in the transverse field measurements. In the superconducting state we observe jump rates which increase from about 0.2 $\mu s^{-1}$ at 12.5 K to about 3 $\mu s^{-1}$ at 40 K just above $T_c$. Even at 40 K the diffusion path of the muons on a microsecond time scale will be shorter than about a nm. Such a distance is negligible as compared to the characteristic length scales of the vortex lattice, the magnetic penetration depth and the vortex-vortex spacing that are of the order of 100 nm and 60 nm (at 0.6 T), respectively.

For an at least qualitative explanation of the muon dynamics one needs to know the muon stopping site within the $MgB_2$ lattice. Since Mg is almost completely ionized in the $MgB_2$ compound, the positive muon will be strongly repelled by the positively charged Mg ions. This should make the boron plane the most likely stopping site. In this plane the site in the center of the boron hexagon has the highest point symmetry $D_{6h}$ (so called b site). The calculated value of the second moment of the field distribution on the b-site gives only $\Delta \sim 0.3$ $\mu s^{-1}$ and thus poorly agrees with the measured value of $\Delta = 0.45$ $\mu s^{-1}$. Much better agreement can be obtained for a muon site, k, that is located about 0.5 Å away from the center of the boron hexagon. Site b is surrounded by sextets of such k-sites (along the center to boron line and, rotated by $\pi/6$, of sites m) which form a ring around the b site. Similar interstitial sites were identified as the muon-sites in the hexagonal compounds $UNi_2Al_3$[30] and $GdNi_5$[31]. There it was argued that the potential barrier between neighboring sites in the ring must be shallow, thus preventing localization of the $\mu^+$ at individual m or k sites even at low temperatures. Hopping of $\mu^+$ along a ring of interstitials as reported recently for GdNi5 [31] may explain our observed low temperature behavior.

We would like to thank Alex Amato and Dierk Herlach (PSI) for the technical support during the μSR experiments. This work was supported by the BMFT. CN gratefully acknowledges the financial support by the Deutsche Forschungsgemeinschaft and helpful discussions with J.I. Budnick and G. Solt.

Figure Captions

Fig. 1: Field dependence of the depolarization rate $\sigma$ at T = 5 K. The observed behavior is characteristic of a pinning induced disorder of the flux line lattice.

Fig.2 : Temperature dependence of $\sigma$ for applied external fields of 50 mT (open squares), 0.3 T (closed circles) and 0.6 T (open circles). Solid lines are the best fits obtained with the function $\sigma(T) = \sigma(0) \cdot (1-(T/T_c)^{\alpha})$. The T-dependence can be seen to change systematically as a function of applied field with the value of $\alpha$ changing from $\alpha$ = 1.33 at 50 mT to $\alpha$ = 1.92 at 0.6 T. In particular, this implies that the T-dependence of $\sigma(T)$ at low T becomes weaker as the applied magnetic field increases.

Fig. 3: Temperature dependence of $\sigma$ for an applied external field of 0.6 T. The solid line shows the best fit using a two-gap model [14]. The dashed line is the result of a $T^2$ fit to the data. Also shown by the dotted line is a fit assuming an isotropic single gap. The inset shows the low temperature region on a larger scale.

Fig. 4: ZF μSR spectra obtained at 1.8 K, 12.5 K, 30 K and 60 K. The Gaussian like decay together with the almost complete recovery of the asymmetry to 1/3 observed at 1.8 K is characteristic for the interaction of the muon spin with randomly oriented static nuclear magnetic moments. The suppression of the 1/3 tail observed in the spectrum at 12.5 K indicates a low mobility of the muons with hopping rates of less than $\Delta^{-1}$. In the limit of fast diffusion ($\Delta\tau_c \ll 1$) the time evolution of the muon polarization changes to an exponential shape such as for the spectrum at 60 K.

Fig.1:

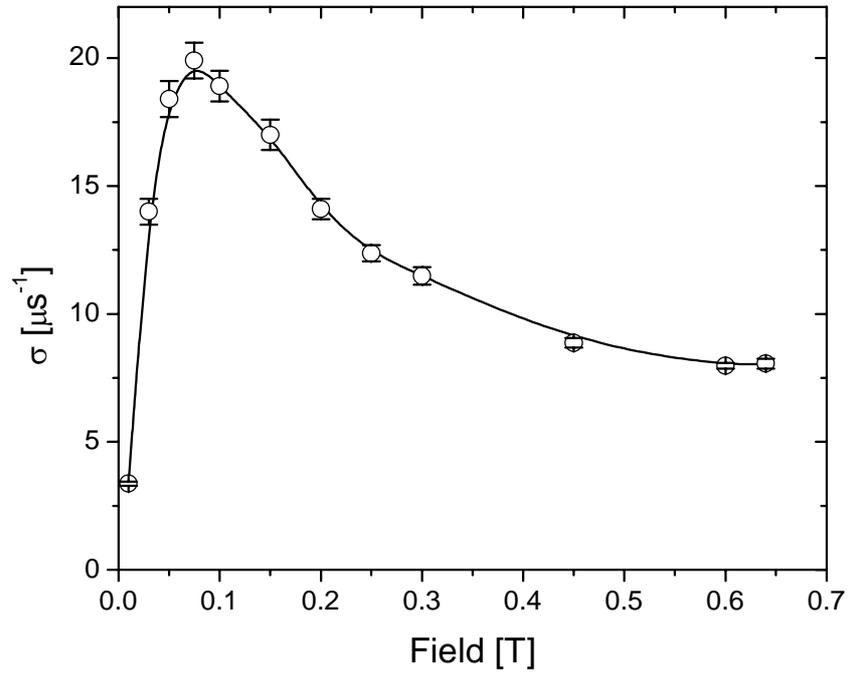

Fig.2:

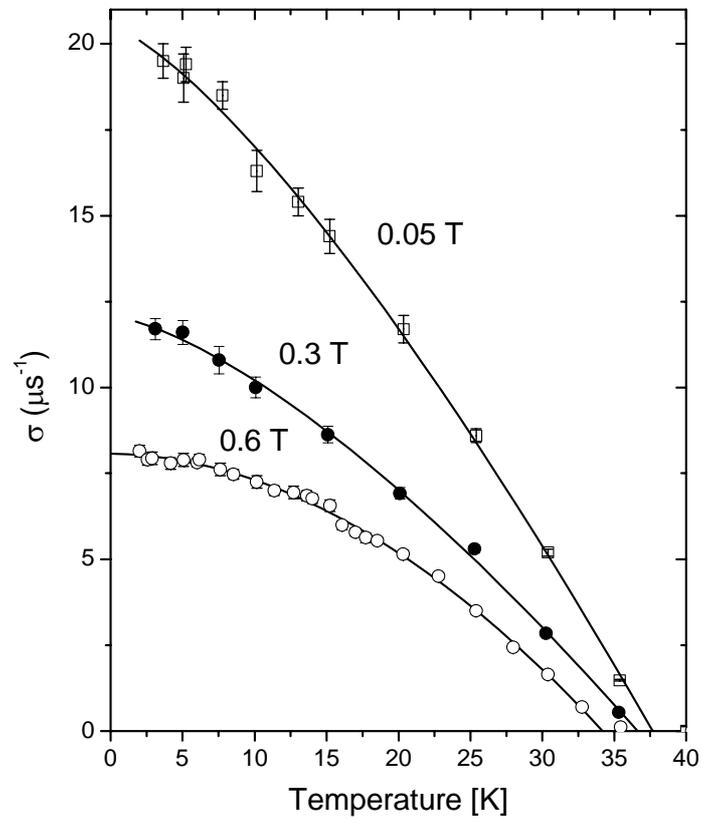

Fig.3:

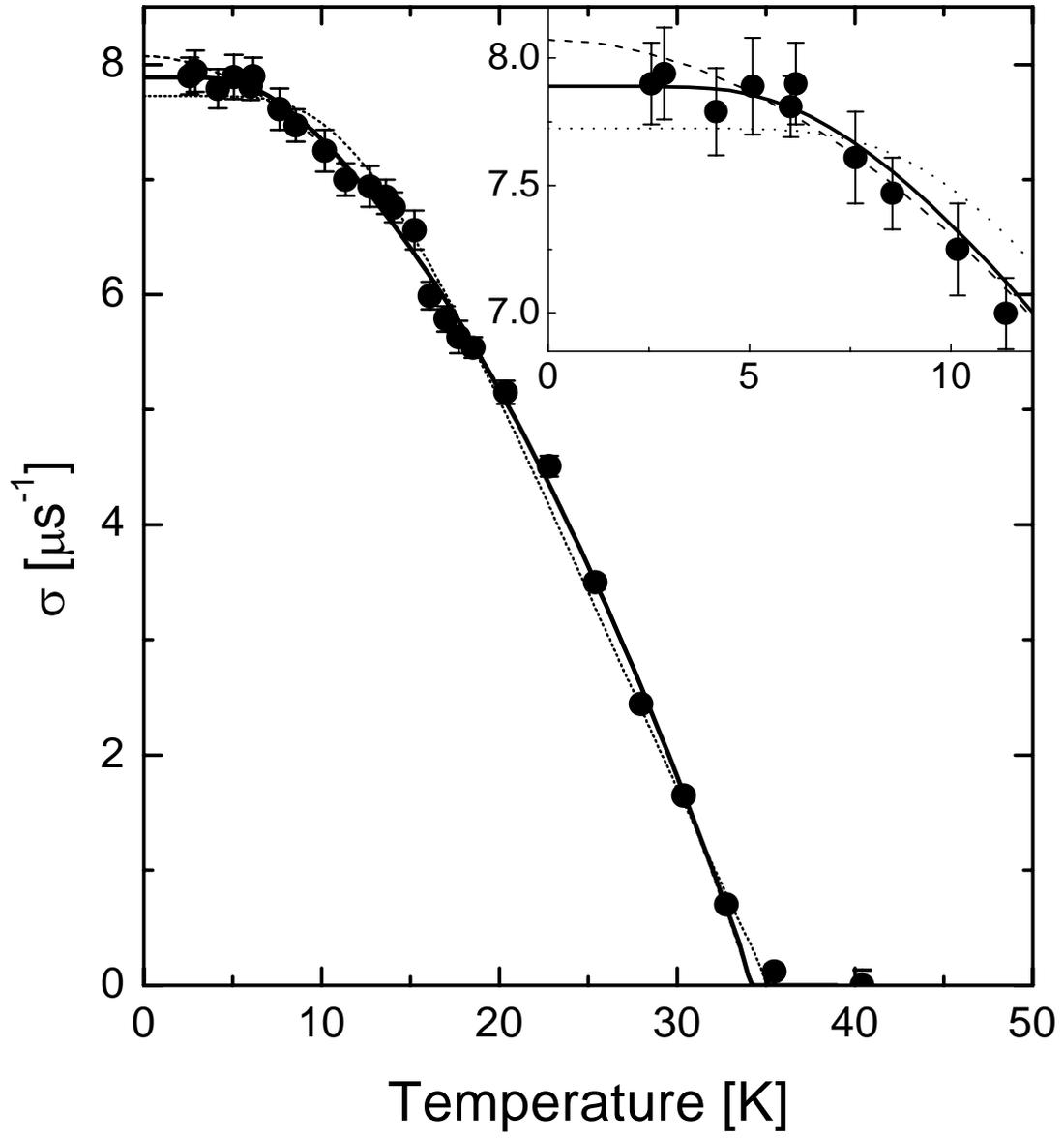

Fig. 4:

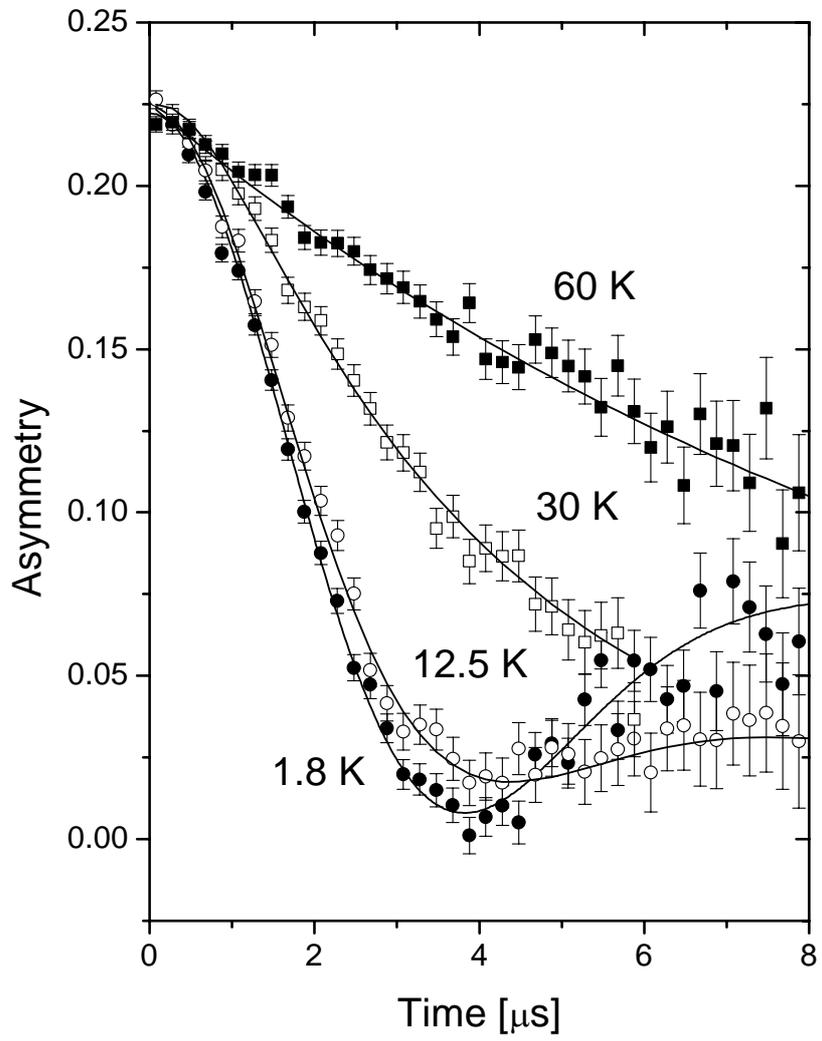

# References

[1] J. Nagamatsu et al., Nature **410**, 63 (2001)
[2] J. Kortus et al., Phys. Rev. Lett. **86**, 4656 (2001)
[3] S. L. Bud`kov, Phys. Rev. Lett. **86**, 1877 (2001)
[4] D. G. Hinks et al., Nature **411**, 457 (2001)
[5] H. Kotegawa et al., cond-mat/0102334
[6] R. Osborn et al., Phys. Rev. Lett. **87**, 017005 (2001); T. Yildirim et al., Phys. Rev. Lett. **87**, 037001 (2001)
[7] Ch. Wälti et al., cond-mat/0102522; R.K. Kremer et al., cond-mat/0102432
[8] T. Takahashi et al., Phys. Rev. Lett. **86**, 4915 (2001)
[9] G. Karapetrov et al., Phys. Rev. Lett. **86**, 4374 (2001)
[10] A.Y. Liu, I.I. Mazin , J. Kortus, cond-mat/0103570
[11] F. Bouquet et al., Phys. Rev. Lett. **87**, 047001 (2001); Y. Wang et al., Physica C **335**, 179 (2001)
[12] X.K. Chen et al., cond-mat/0104005v3
[13] C. Panagopoulos et al., Phys. Rev. **B64**, 094514 (2001)
[14] F. Bouquet et al., cond-mat/0107196
[15] M. Xu et al. cond-mat/0105271, S. Lee et al. cond-mat/0105545
[16] S. Patnaik et al., cond-mat/0104562
[17] O. F. de Lima et al., Phys. Rev. Lett. **86**, 5974 (2001)
[18] W Bardford and J.M.F Gunn, Physica **C156**, 515 (1988)
[19] E.H. Brandt, Phys. Rev. **B37**, 2349 (1988)
[20] B. Pümpin et al., Phys. Rev. **B42**, 8019 (1990)
[21] Ch. Niedermayer et al., Journal of Superconductivity **7**, 165 (1994)
[22] B. Mühlschlegel, Z. Physik, **155**, 313 (1959)
[23] F. Manzano and A. Carrington, cond-mat/0106166
[24] S. Haas and K. Maki, cond-mat/0104207
[25] T. Schneider and M. Frick in Earlier and recent Aspects of Superconductivity, Vol. 90 of Springer Series in Solid State Sciences, 501-517
[26] P.J. Hirschfeld and N. Goldenfeld, Phys. Rev. B **48**, 4219 (1993)
[27] D. Harshman et al., Phys. Rev. B **39**, 851 (1989)
[28] W.N. Hardy et al., Phys. Rev. Lett. **70**, 3999 (1993)
[29] R. S. Hayano et al., Phys. Rev. B **20**, 850 (1979)
[30] A. Amato et al., Physica B **289-290** (2000) 447
[31] A.M. Mulders et al., Physica B **289-290** (2000) 451